\documentclass[aps,prl,twocolumn,superscriptaddress]{revtex4-2}

\usepackage{graphicx}
\usepackage{dcolumn}
\usepackage{bm}
\usepackage{amssymb}
\usepackage{mathrsfs}
\usepackage{amsmath}
\usepackage{comment}
\usepackage{threeparttable}

\usepackage{newtxtext}

\PassOptionsToPackage{breaklinks}{hyperref}
\usepackage{hyperref}
\hypersetup{	colorlinks=true,
			linkcolor=blue,
			citecolor=blue,
			urlcolor=blue,
		}	
\usepackage{url}
\usepackage[hyphenbreaks]{breakurl}
\usepackage{graphicx}

\def\klpionn{K_L\!\to\!\pi^0\nu\overline{\nu}}

\def\klpioxo{K_L\!\to\!\pi^0X^0}
\def\klpiopiopio{K_L\!\to\!3\pi^0}
\def\klpipipio{K_L\!\to\!\pi^+\pi^-\pi^0}
\def\klpiopio{K_L\!\to\!2\pi^0}
\def\klgg{K_L\!\to\!2\gamma}
\def\klpienu{K_L\!\to\!\pi^{\pm}e^{\mp}\nu}
\def\kcpienu{K^{\pm}\!\to\!\pi^0e^{\pm}\nu}

\def\kcpipio{K^{\pm}\!\to\!\pi^{\pm}\pi^0}
\def\to{\rightarrow}
\def\Zvtx{{Z_{\mathrm{vtx}}}}
\def\Pt{P_{t}}
\def\pmev{\mathrm{MeV}/c}

\begin{document}

\title{ Study of the $\klpionn$ Decay at the J-PARC KOTO Experiment }


\newcommand{\InstKorea}{\affiliation{Department of Physics, Korea University, Seoul 02841, Republic of Korea}}
\newcommand{\InstOsaka}{\affiliation{Department of Physics, Osaka University, Toyonaka, Osaka 560-0043, Japan}}
\newcommand{\InstMichigan}{\affiliation{Department of Physics, University of Michigan, Ann Arbor, Michigan 48109, USA}}
\newcommand{\InstChicago}{\affiliation{Enrico Fermi Institute, University of Chicago, Chicago, Illinois 60637, USA}}
\newcommand{\InstNTU}{\affiliation{Department of Physics, National Taiwan University, Taipei, Taiwan 10617, Republic of China}}
\newcommand{\InstArizona}{\affiliation{Department of Physics, Arizona State University, Tempe, Arizona 85287, USA}}
\newcommand{\InstSaga}{\affiliation{Department of Physics, Saga University, Saga 840-8502, Japan}}
\newcommand{\InstKyoto}{\affiliation{Department of Physics, Kyoto University, Kyoto 606-8502, Japan}}
\newcommand{\InstKEK}{\affiliation{Institute of Particle and Nuclear Studies, High Energy Accelerator Research Organization (KEK), Tsukuba, Ibaraki 305-0801, Japan}}
\newcommand{\InstYamagata}{\affiliation{Department of Physics, Yamagata University, Yamagata 990-8560, Japan}}
\newcommand{\InstJeonbuk}{\affiliation{Division of Science Education, Jeonbuk National University, Jeonju 54896, Republic of Korea}}
\newcommand{\InstJPARC}{\affiliation{J-PARC Center, Tokai, Ibaraki 319-1195, Japan}}
\newcommand{\InstNDA}{\affiliation{Department of Applied Physics, National Defense Academy, Kanagawa 239-8686, Japan}}
\newcommand{\InstOkayama}{\affiliation{Research Institute for Interdisciplinary Science, Okayama University, Okayama 700-8530, Japan}}
\InstKorea
\InstMichigan
\InstNTU
\InstArizona
\InstChicago
\InstOsaka
\InstKEK
\InstKyoto
\InstJeonbuk
\InstJPARC
\InstOkayama
\InstNDA
\InstYamagata
\InstSaga
\author{J.~K.~Ahn}\InstKorea
\author{B.~Beckford}\InstMichigan
\author{M.~Campbell}\InstMichigan
\author{S.~H.~Chen}\InstNTU
\author{J.~Comfort}\InstArizona
\author{K.~Dona}\InstMichigan
\author{M.~S.~Farrington}\InstChicago
\author{K.~Hanai}\InstOsaka
\author{N.~Hara}\InstOsaka
\author{H.~Haraguchi}\InstOsaka
\author{Y.~B.~Hsiung}\InstNTU
\author{M.~Hutcheson}\InstMichigan
\author{T.~Inagaki}\InstKEK
\author{M.~Isoe}\InstOsaka
\author{I.~Kamiji}\InstKyoto
\author{T.~Kato}\InstOsaka
\author{E.~J.~Kim}\InstJeonbuk
\author{J.~L.~Kim}\InstJeonbuk
\author{H.~M.~Kim}\InstJeonbuk
\author{T.~K.~Komatsubara}\InstKEK\InstJPARC
\author{K.~Kotera}\InstOsaka
\author{S.~K.~Lee}\InstJeonbuk
\author{J.~W.~Lee}\thanks{Present address: Department of Physics, Korea University, Seoul 02841, Republic of Korea.}\InstOsaka
\author{G.~Y.~Lim}\InstKEK\InstJPARC
\author{Q.~S.~Lin}\InstChicago
\author{C.~Lin}\InstNTU
\author{Y.~Luo}\InstChicago
\author{T.~Mari}\InstOsaka
\author{T.~Masuda}\InstOkayama
\author{T.~Matsumura}\InstNDA
\author{D.~Mcfarland}\InstArizona
\author{N.~McNeal}\InstMichigan
\author{K.~Miyazaki}\InstOsaka
\author{R.~Murayama}\thanks{Present address: RIKEN Cluster for Pioneering Research, RIKEN, Wako, Saitama 351-0198 Japan.}\InstOsaka
\author{K.~Nakagiri}\thanks{Present address: Department of Physics, University of Tokyo, Tokyo 113-0033, Japan.}\InstKyoto
\author{H.~Nanjo}\thanks{Present address: Department of Physics, Osaka University, Toyonaka, Osaka 560-0043, Japan.}\InstKyoto
\author{H.~Nishimiya}\InstOsaka
\author{Y.~Noichi}\InstOsaka
\author{T.~Nomura}\InstKEK\InstJPARC
\author{T.~Nunes}\InstOsaka
\author{M.~Ohsugi}\InstOsaka
\author{H.~Okuno}\InstKEK
\author{J.~C.~Redeker}\InstChicago
\author{J.~Sanchez}\InstMichigan
\author{M.~Sasaki}\InstYamagata
\author{N.~Sasao}\InstOkayama
\author{T.~Sato}\InstKEK
\author{K.~Sato}\thanks{Present address: Institute for Space-Earth Environmental Research, Nagoya University, Nagoya, Aichi 464-8601, Japan.}\InstOsaka
\author{Y.~Sato}\InstOsaka
\author{N.~Shimizu}\InstOsaka
\author{T.~Shimogawa}\thanks{Present address: KEK, Tsukuba, Ibaraki 305-0801, Japan.}\InstSaga
\author{T.~Shinkawa}\InstNDA
\author{S.~Shinohara}\thanks{Present address: Department of Physics, Osaka University, Toyonaka, Osaka 560-0043, Japan.}\InstKyoto
\author{K.~Shiomi}\InstKEK\InstJPARC
\author{R.~Shiraishi}\InstOsaka
\author{S.~Su}\InstMichigan
\author{Y.~Sugiyama}\thanks{Present address: KEK, Tsukuba, Ibaraki 305-0801, Japan.}\InstOsaka
\author{S.~Suzuki}\InstSaga
\author{Y.~Tajima}\InstYamagata
\author{M.~Taylor}\InstMichigan
\author{M.~Tecchio}\InstMichigan
\author{M.~Togawa}\thanks{Present address: KEK, Tsukuba, Ibaraki 305-0801, Japan.}\InstOsaka
\author{T.~Toyoda}\InstOsaka
\author{Y.-C.~Tung}\thanks{Present address: Department of Physics, National Taiwan University, Taipei, Taiwan 10617, Republic of China}\InstChicago
\author{Q.~H.~Vuong}\InstOsaka
\author{Y.~W.~Wah}\InstChicago
\author{H.~Watanabe}\InstKEK\InstJPARC
\author{T.~Yamanaka}\InstOsaka
\author{H.~Y.~Yoshida}\InstYamagata
\author{L.~Zaidenberg}\InstMichigan
\collaboration{KOTO Collaboration} \noaffiliation

\begin{abstract}
	The rare decay $\klpionn$ was studied with the dataset taken at the J-PARC KOTO experiment in 2016, 2017, and 2018. 
	With a single event sensitivity of $( 7.20 \pm 0.05_{\rm stat} \pm 0.66_{\rm syst} ) \times 10^{-10}$, 
	three candidate events were observed in the signal region. 
	After unveiling them, contaminations from $K^{\pm}$ and scattered $K_L$ decays were studied, 
	and the total number of background events was estimated to be $1.22 \pm 0.26$.  
	We conclude that the number of observed events is statistically consistent with the background expectation. 
	For this dataset, we set an upper limit of $4.9 \times 10^{-9}$ on the branching fraction of $\klpionn$ at the 90\% confidence level.
\end{abstract}


\maketitle

\paragraph*{Introduction.---}\hspace{-14pt}
The rare kaon decay $\klpionn$ directly breaks $CP$ symmetry \cite{Littenberg,Kaon_Review} and has a highly suppressed branching fraction predicted to be $(3.00 \pm 0.30) \times 10^{-11}$ in the standard model (SM) \cite{KLpi0nunuSM}. 
The accurate prediction of the branching fraction makes this decay sensitive to new physics beyond the SM (e.g.~\cite{Kpinunu_BSM_2016:1,Kpinunu_BSM_2016:2}). 
The current best upper limit on the branching fraction is $3.0 \times 10^{-9}$ at the 90\% confidence level (C.L.) \cite{KOTO2015} set by the KOTO experiment \cite{KOTOproposal,KOTO} at the Japan Proton Accelerator Research Complex (J-PARC) \cite{J-PARC} with the dataset taken in 2015. 
An indirect upper limit, called the Grossman-Nir bound \cite{GNlimit}, of $7.8 \times 10^{-10}$ is set using the $K^+\rightarrow\pi^+\nu\bar{\nu}$ decay \cite{NA62_2017}. 

The KOTO experiment is dedicated to studying the $\klpionn$ decay. 
We presented preliminary findings on the $\klpionn$ search based on data accumulated from 2016 to 2018 at a conference \cite{KAON2019_Shinohara}. 
At the time, we reported the observation of four candidate events in the signal region with a small background expectation. 
In this Letter, we conclude our findings with the 2016--2018 dataset after reanalyzing the data and studying additional sources of background contamination. 
Note that KOTO is also sensitive to the $\klpioxo$ decay (e.g. \cite{KLpi0X0:1,KLpi0X0:2,KLpi0X0:3,KLpi0X0:4}), where $X^0$ is an invisible light boson, but this Letter focuses on the analysis of the $\klpionn$ search. 

\paragraph*{Experimental methods and apparatus.---}\hspace{-12pt}
\begin{figure*}
	\includegraphics[width=1\linewidth]{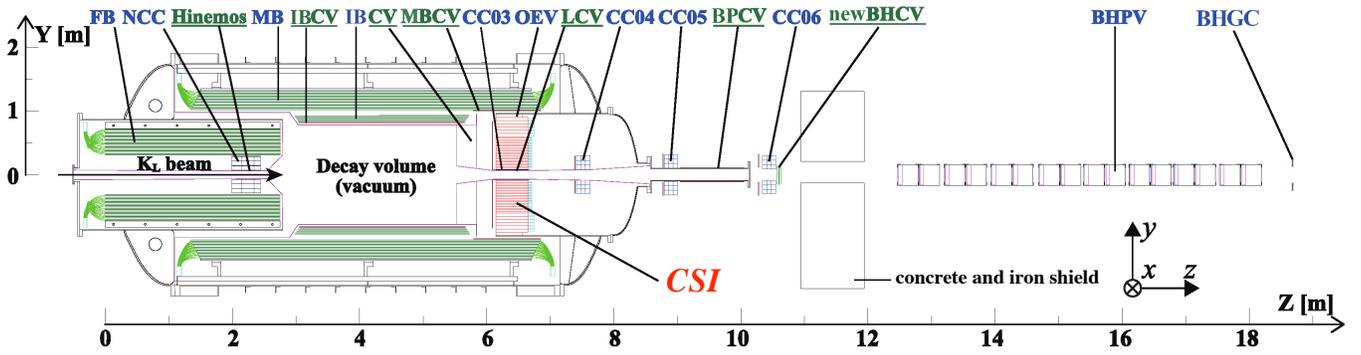}
	\caption{ 	Cross sectional view of the KOTO detector.
			The beam enters from the left.
			Detector components with their abbreviated names written in blue (in green and underlined) are photon (charged-particle) veto counters; see text for information about the abbreviations.
		}
	\label{fig:KOTODetector}
\end{figure*}
A 30-GeV proton beam from the J-PARC main ring is incident on a gold production target \cite{J-PARC_HEF_AuTarget} in the Hadron Experimental Facility. 
Particles produced at the target are guided through a 20-m-long beam line consisting of two collimators and a sweeping magnet located between them \cite{KOTO_BeamLine}. 
At the end of the downstream collimator (beam exit), the beam consists of neutrons, photons, and $K_L$'s and has a size of $8\times8$ cm$^2$. 
The peak $K_L$ momentum is 1.4 GeV/$c$ \cite{PhD_Sato}. The $K_L$ flux was measured to be $2.1 \times 10^{-7}$ $K_L$'s per proton on target (POT) \cite{PhD_Nakagiri}. 
Beam particles that leak outside the nominal beam size due to interactions with beam line components are referred to as ``beam-halo" particles. 

The cross sectional view of the KOTO detector is shown in Fig.~\ref{fig:KOTODetector}. 
The origin of the $z$ axis, which lies along the beam direction, is the upstream edge of the KOTO detector, 21.5 m away from the target. 
The $x$ (horizontal) and $y$ (vertical) axes are defined using a right-handed coordinate system. 
The detectable particles in the final state of $\klpionn$ are the two photons from the $\pi^{0}$ decay. 
We measure the photon energy and timing with a 2-m-diameter cylindrical electromagnetic calorimeter (CSI) \cite{KOTOdet_CSI} centered along the beam axis with a $15\times15$ cm$^2$ beam hole. 
The CSI is composed of 2716 undoped-CsI crystals that have a length of 50 cm and a cross section of $2.5 \times 2.5 \ {\rm cm^2}$ ($5 \times 5 \ {\rm cm^2}$) inside (outside) the central $1.2 \times 1.2 \ {\rm m^2}$ region. 
To ensure that there are no other detectable particles, the decay volume is surrounded with hermetic veto counters. 
Photon veto counters consist of undoped-CsI crystal counters (NCC, OEV, CC03, CC04, CC05, and CC06), or lead-scintillator (FB, MB, and IB), lead-aerogel (BHPV), and lead-acrylic (BHGC) counters; charged-particle veto counters are made of plastic scintillators (Hinemos, IBCV, MBCV, CV, LCV, and BPCV) or wire chambers (newBHCV).
A cylindrical photon veto counter named Inner Barrel (IB) \cite{KOTOdet_IB} was installed and used since 2016. 
The waveform from the detectors is recorded by either 125-MHz \cite{KOTO_125MHzFADC} or 500-MHz sampling ADCs \cite{KOTO_500MHzFADC}.
Details of the apparatus are available in \cite{KOTO2015}. 

\paragraph*{Data taking.---}\hspace{-14pt}
The data taken in 2016--2018 corresponds to $3.05 \times 10^{19}$ protons on target. 
In this dataset, the primary proton beam was extracted every 5.2--5.5 sec from the J-PARC main ring for a duration of 2 sec.
The beam power varied from 31 to 51 kW. 
The $K_L$ incident rate on the KOTO detector varied from 4 to 7 MHz. 
Physics triggers which were organized around a first-level trigger (L1) and a second-level trigger (L2) were used in the data acquisition to collect the $\klpionn$ signal sample. 
L1 required the total deposited energy in CSI to be larger than 550 MeV with no coincident hit in NCC, MB, IB, CV, and CC03. 
In 2018, L1 further required no coincident hit in CC04, CC05, and CC06. 
In 2016, L2 calculated the position of the center of deposited energy (COE) in CSI, defined as ${\bm R}_{\mathrm{COE}} = \sum {e_i {\bm r}_i} / \sum {e_i}$ where $e_i$ and ${\bm r}_i$ are the deposited energy and the $(x,y)$ position of each CSI crystal, respectively, and selected events whose $R_{\mathrm{COE}}$ was larger than 165 mm \cite{DAQTrigger_2013run}. 
In 2017 and 2018, L2 counted the number of electromagnetic showers in CSI and selected the events with the desired number of showers \cite{KAON2019_Jay}. 
The number of triggered events in physics triggers was $6.55 \times 10^9$. 
$\klpiopio$, $\klpiopiopio$, and $\klgg$ decay samples were collected with another trigger using only L1 with a prescale factor. 

\paragraph*{Event reconstruction and selection.---}\hspace{-11pt}
In the off-line analysis, adjacent crystals with deposited energies larger than 3 MeV in CSI were grouped into a cluster, which was used to reconstruct the photon energy, timing, and position. 
The opening angle ($\theta$) between the two photons was calculated from $\cos\theta=1- M^2_{\pi^0} / (2E_{\gamma_1}E_{\gamma_2})$, where $M_{\pi^0}$ is the nominal $\pi^0$ mass, and $E_{\gamma_1}$ and $E_{\gamma_2} \ (\!<\! E_{\gamma_1}$) are the energies of the two photons. 
Using the opening angle and assuming the $\pi^0 \to 2 \gamma$ decays on the beam axis, the $\pi^{0}$ decay vertex position ($\Zvtx$) and the $\pi^0$ four momentum were calculated. 

The $\pi^{0}$ from $\klpionn$ decays is expected to have a finite transverse momentum ($\Pt$) due to the neutrinos. 
We defined the signal region in the $\Pt$ and $\Zvtx$ plane as the area encompassing $130 \!<\! \Pt \!<\! 250 \ \pmev$ and $3200 \!<\! \Zvtx \!<\! 5000 \ {\rm mm}$ excluding the area with $\Pt \!<\! 1/35 (\Zvtx - 4000 \ {\rm mm})+130$ $\pmev$ for $4000 \!<\! \Zvtx \!<\! 5000 \ {\rm mm}$ in order to suppress the background from $\klpipipio$ decays. 

To avoid bias, the event selection criteria (cuts) were determined using data collected outside the blind region defined by $120 \!<\! \Pt \!<\! 260 \ \pmev$ and $2900 \!<\! \Zvtx \!<\! 5100 \ {\rm mm}$. 
The selected events were required to have $(E_{{\gamma}_1} + E_{{\gamma}_2}) >\! 650$ MeV and $R_{\mathrm{COE}} >\! 200$ mm in CSI to avoid trigger inefficiency. 
The photon energy was required to be $100 \!< E_{\gamma} <\! 2000$ MeV and the photon position $(x,y)$ was required to be within the CSI fiducial region defined as $\sqrt{x^2 + y^2}\!<\!850\ {\rm mm}$ and $\min(|x|,|y|)\!>\!150\ {\rm mm}$. 
The timing difference between the two photons was required to be within 1 ns, and their distance to be larger than $300\ {\rm mm}$ to ensure cluster separation. 
The ratio between the energies of the two photons ($E_{\mathrm{ratio}}$), defined as $E_{\mathrm{ratio}} = E_{\gamma_2} / E_{\gamma_1}$, was required to be larger than 0.2. 
The product of the photon energy and the photon momentum angle with reference to the beam axis ($E\theta$) was required to be larger than 2500 MeV$\cdot$deg; the requirement on $E_{\mathrm{ratio}}$ and $E\theta$ reduces the $\klpiopio$ background from photon miscombinations in the $\pi^0$ reconstruction.
The opening angle between the photon directions projected on the $x$-$y$ plane (projection angle) was required to be less than $150^\circ$ to reduce the $\klgg$ background. 
Events were discarded if a veto counter had a hit with a deposited energy above its given threshold and a timing within its given veto window. 
Finally, shape-related cuts based on each cluster in CSI and the waveform of each CSI crystal, described later, were used to reduce background events from neutrons. 

\begin{figure}
	\begin{center}
		\includegraphics[width=1\linewidth]{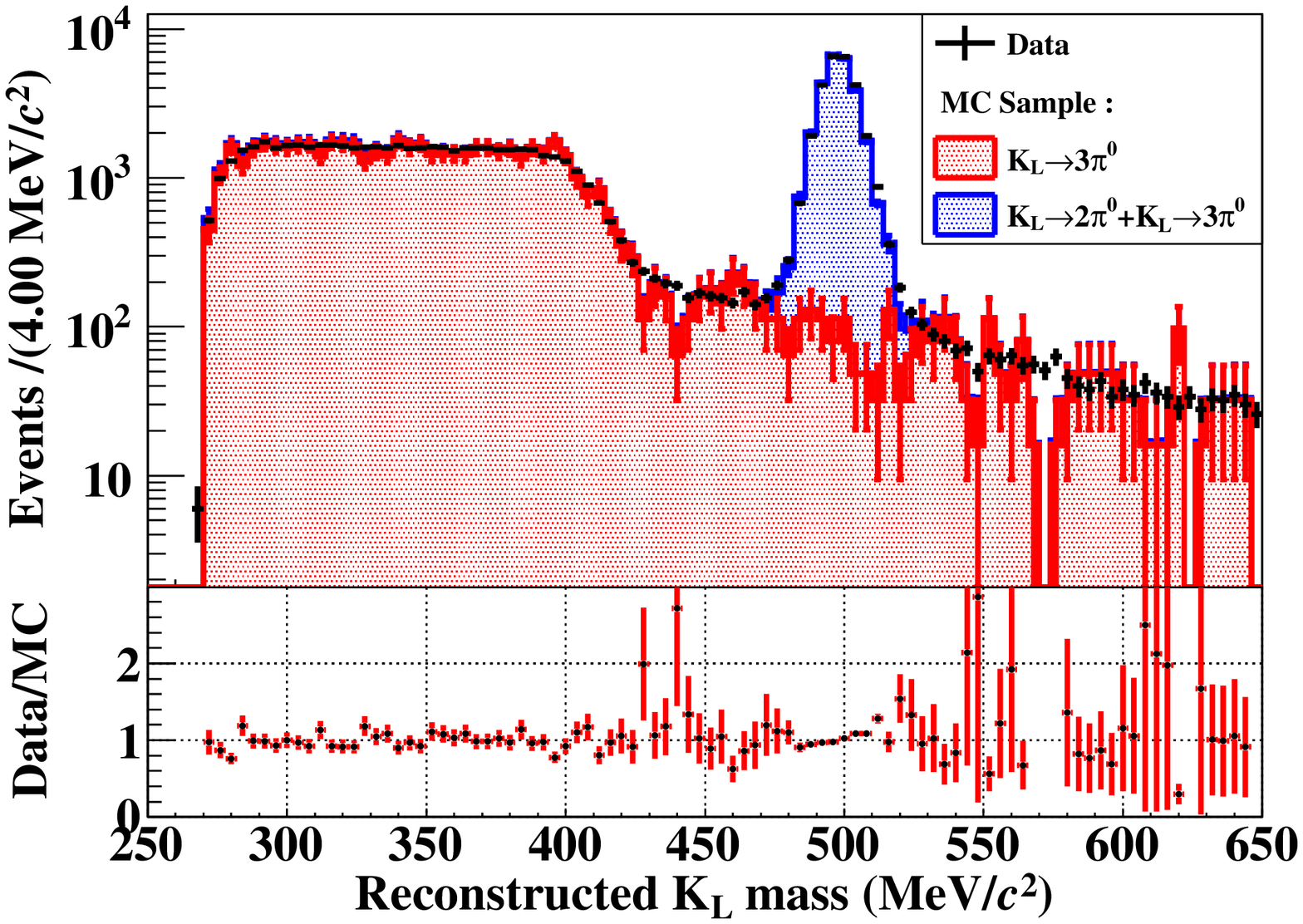}
		\caption{	
				Reconstructed $K_L$ mass ($M_{K_L}$) distribution of $\klpiopio$ events after imposing the $\klpiopio$ selection criteria except for the requirement on $M_{K_L}$. 
				The bottom panel shows the ratio of data and MC calculations for each histogram bin.
				}
		\label{fig:KLmass}
	\end{center}
\end{figure}

\paragraph*{Normalization and single event sensitivity.---}\hspace{-12pt}
The acceptance for $\klpionn$ ($A_{\rm sig}$) was evaluated using \mbox{\textsc{Geant4}}-based \cite{GEANT4:1,GEANT4:2,GEANT4:3} Monte Carlo (MC) simulations. 
Accidental activity in detectors was recorded with a random trigger during physics data taking and was overlaid on the MC events.
The single event sensitivity (SES) was normalized with the $\klpiopio$ decay sample. 
To reconstruct $\klpiopio$ decays, we used events with four clusters in CSI. 
Among the three possible $\pi^0$ pairs from photon combinatorics, we selected the one with the smallest $\Zvtx$ difference. 
We required the reconstructed invariant $K_L$ mass ($M_{K_L}$) to be within $\pm 15$ MeV/$c^2$ of the nominal $K_L$ mass. 
Figure~\ref{fig:KLmass} shows the $M_{K_L}$ distribution after imposing the kinematic cuts for $\klpiopio$ and the veto cuts except for the requirement on $M_{K_L}$. 
The SES was obtained as:
\begin{equation}
	{\rm SES} = \frac{1}{A_{\rm sig}} \frac{A_{\rm norm}\ {\rm Br}(\klpiopio)}{N_{\rm norm}},
	\label{eq:SES}
\end{equation}
where $A_{\rm norm}$ is the acceptance of $\klpiopio$ decays after taking into account other $K_L$ decay contaminations, 
${\rm Br}(\klpiopio)$ is the branching fraction of $\klpiopio$ \cite{PDG2020}, 
and $N_{\rm norm}$  is the number of events after imposing the $\klpiopio$ selection criteria with a correction of the prescale factor applied to the trigger. 
Based on $A_{\rm sig} = 0.61$\%, $A_{\rm norm} = 0.41$\%, and $N_{\rm norm} = 7.98 \times 10^5$, the SES was estimated to be $(7.20 \pm 0.05_{\rm stat} \pm 0.66_{\rm syst})\times 10^{-10}$, 
which corresponded to 1.8 times better sensitivity than the previous search \cite{KOTO2015}. 

The systematic uncertainties on the SES are summarized in Table~\ref{tab:Systematics}. 
The two largest uncertainties were from shape-related cuts and inconsistencies in the normalization procedure. 
The discrepancy in the acceptance between data and MC simulations for each shape-related cut was studied using a sample of $\pi^0$'s from the reconstructed $\klpiopio$ events, and their sum in quadrature was calculated to be 5.2\%. 
The uncertainty on the normalization was calculated as the maximum difference between the SES calculated using $\klpiopio$, $\klpiopiopio$, and $\klgg$ decays, and was estimated to be 5.2\%. 
\begin{table}
\caption{Summary of relative systematic uncertainties on the single event sensitivity.}
\label{tab:Systematics}
\centering
\begin{tabular}{lD{.}{.}{-1}}
	\hline \hline
	source & \multicolumn{1}{c} ~uncertainty [\%]\\
	\hline
	trigger effect & 0.26\\
	photon selection cuts & 0.57\\
	kinematic cuts for $\klpionn$ & 2.9\\
	veto cuts & 3.2\\
	shape-related cuts & 5.2\\
	$K_L$ momentum spectrum & 0.98\\
	kinematic cuts for $\klpiopio$ & 3.2\\
	$\klpiopio$ branching fraction & 0.69\\
	normalization modes inconsistency & 5.2\\
	\hline
	total &  9.2\\
	\hline \hline
\end{tabular}
\end{table}

\paragraph*{Background estimation.---}\hspace{-12pt}
Table~\ref{tab:BGSummary} summarizes the expected numbers of background events for which we calculated each central value and its uncertainty. 
The total number of background events in the signal region was estimated to be $1.22 \pm 0.26$ by adding the central values of each background source. 
Note that the backgrounds from $K^{\pm}$ and beam-halo $\klgg$ decays were not known when we first looked inside the blind region, and they were added after performing the studies described later in this Letter. 
\begin{table}
	\caption{Summary of the numbers of background events with a central value estimate.}
	\label{tab:BGSummary}
	\centering
	\begin{threeparttable}[h]
	\begin{tabular}{llc}
		\hline \hline
		source & & Number of events\\
		\hline
		$K_L$			& $\klpiopiopio$					& 0.01 $\pm$ 0.01 \\
						& $\klgg$	(beam halo)				& 0.26 $\pm$ 0.07 \tnote{a}\\
						& Other $K_L$ decays 				& 0.005 $\pm$ 0.005 \\
		$K^{\pm}$ 		& 								& 0.87 $\pm$ 0.25 \tnote{a}\\
		Neutron			& Hadron cluster					& 0.017 $\pm$ 0.002\\
						& CV $\eta$ 						& 0.03 $\pm$ 0.01\\
						& Upstream $\pi^0$ 					& 0.03 $\pm$ 0.03\\
		\hline
		total 				&								&1.22 $\pm$ 0.26\\
		\hline \hline
	\end{tabular}
	\begin{tablenotes}
		\item[a] Background sources studied after looking inside the blind region. 
	\end{tablenotes}
	\end{threeparttable}
\end{table}

The $\klpiopiopio$ background arises from photon detection inefficiency in veto counters mainly due to accidental hits overlapping a photon pulse and shifting its measured time outside the veto window. 
To suppress this type of background, a pulse-shape discrimination method was introduced by applying a fast Fourier transform (FFT) to the waveform recorded by the veto counters. 
We prepared templates in the frequency domain of the single hit waveform collected from data, and calculated a $\chi^2$ value based on the difference between the observed waveform and the template. 
When the $\chi^2$ value exceeded a given threshold, the veto window was widened to accommodate possible timing shifts due to overlapping pulses. 
The number of background events from $K_L\rightarrow3\pi^0$ was studied with the MC simulation, and estimated to be $0.01 \pm 0.01$. 
The numbers of $\klpiopio$, $\klpipipio$, and $\klpienu$ background events were estimated to be $\!<\! 0.08$, $\!<\! 0.02$, and $\!<\! 0.08$ (90\% C.L.), respectively.
Backgrounds from other $K_L$ decays were estimated using MC simulations and their aggregate number was estimated to be $0.005 \pm 0.005$. 

The hadron-cluster background is caused by two hadronic clusters being misidentified as photon clusters in CSI. 
This can occur when a beam-halo neutron hits the CSI and produces a cluster, and another neutron from the hadronic interaction produces an additional cluster. 
With the insertion of a 10-mm-thick aluminum plate in the beam at $Z = - 634$ mm, we collected a control sample with an enhanced number of scattered neutrons hitting the CSI. 
Using this sample, an algorithm using a convolution neural network was developed to discriminate neutrons from photons, based on their cluster's energy and timing patterns in CSI as well as their reconstructed incident angle. 
Additional discrimination power was obtained by applying the FFT to the waveform of each CSI crystal and calculating the likelihood ratio of templates in the frequency domain for both the photon and neutron clusters. 
The combined reduction of these shape-related cuts against hadron-cluster events ($R_{\rm shape}$) was estimated to be $(1.8\pm0.2) \times 10^{-6}$ after taking into account photon contaminations in the control sample. 
The number of background events was calculated from $R_{\rm shape} \times \alpha \times N_{\rm Al}$ and was estimated to be $0.017\pm0.002$, 
where $\alpha$ is the ratio of the number of signal and control sample events in the region of $120 \!<\! \Pt \!<\! 500 \ \pmev$ and $2900 \!<\! \Zvtx \!<\! 6000 \ {\rm mm}$ excluding the blind region before imposing shape-related cuts, 
and $N_{\rm Al}$ is the number of control sample events in the signal region before imposing shape-related cuts. 

The CV-$\eta$ and CV-$\pi^0$ backgrounds are generated when beam-halo neutrons hit CV \cite{KOTOdet_CV} and produce $\eta$ and $\pi^0$, respectively.
CV is a charged-particle veto counter made of plastic scintillator strips and located in front of CSI. 
The upstream-$\pi^0$ background is generated when beam-halo neutrons hit NCC and produce $\pi^0$. NCC is located upstream of the decay volume. 
These backgrounds were studied with MC simulations, and the yields were normalized with the ratio between data and MC for events in the region of $\Zvtx \!>\! 5100 \ {\rm mm}$ for the CV-$\eta$ and CV-$\pi^0$ background and the region of $\Zvtx \!<\! 2900 \ {\rm mm}$ for the upstream-$\pi^0$ background with loose selection criteria. 
The numbers of CV-$\eta$, CV-$\pi^0$, and upstream-$\pi^0$ background events were estimated to be $0.03 \pm 0.01$, $\!<\! 0.10$ (90\% C.L.), and $0.03 \pm 0.03$, respectively.

\begin{figure}
	\begin{center}
		\includegraphics[width=1\linewidth]{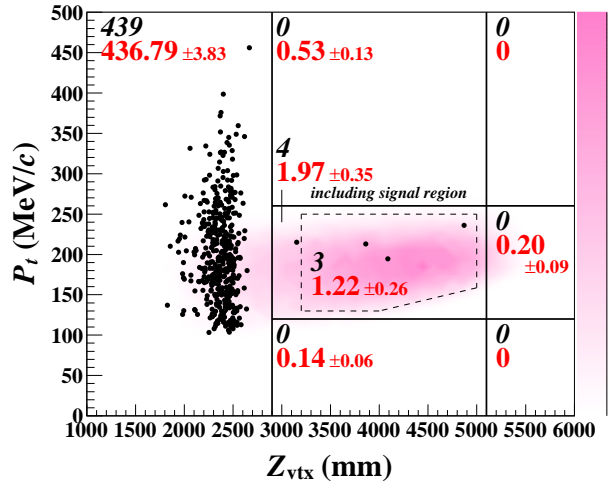}
		\caption{	
				Reconstructed $\pi^0$ transverse momentum ($P_t$) versus $\pi^0$ decay vertex position ($Z_{\rm vtx}$) plot of the events after imposing the $\klpionn$ selection criteria. 
				The region surrounded by dotted lines is the signal region. 
				The black dots represent observed events, and the shaded contour indicates the $\klpionn$ distribution from the MC simulation. 
				The black italic (red regular) numbers indicate the number of observed (background) events for different regions. 
				In particular, $1.22 \pm 0.26 \ (1.97 \pm 0.35)$ is the background expectation for the three (four) events observed inside the signal (blind) region.	
			}
		\label{fig:FinalPtZ}
	\end{center}
\end{figure}

\paragraph*{Examining the blind region.---}\hspace{-12pt}
With the background estimation excluding $K^{\pm}$ and beam-halo $\klgg$ decays, we proceeded to unblind the analysis and observed four candidate events in the signal region and one extra event in the blind region \cite{KAON2019_Shinohara}. 
After we found an incorrect parameter setting which affects the timing used to veto events with multiple pulses in the veto counters, the data were processed again. 
After imposing the same selection criteria to this sample, three of the original four candidate events in the signal region remained as shown in Fig.~\ref{fig:FinalPtZ}. 
Of these, the second event from the right in Fig.~\ref{fig:FinalPtZ} has overlapped pulses in NCC.
The probability of observing such an event is 2.2\%. 
The other events in the blind region have no such features. 

\paragraph*{Background studies after examining the blind region.---}\hspace{-12pt}
Two new types of backgrounds, one from $K^{\pm}$ decays and one from beam-halo $\klgg$ decays, were found and studied after examining the blind region. 

\begin{figure}
	\begin{center}
		\includegraphics[width=1\linewidth]{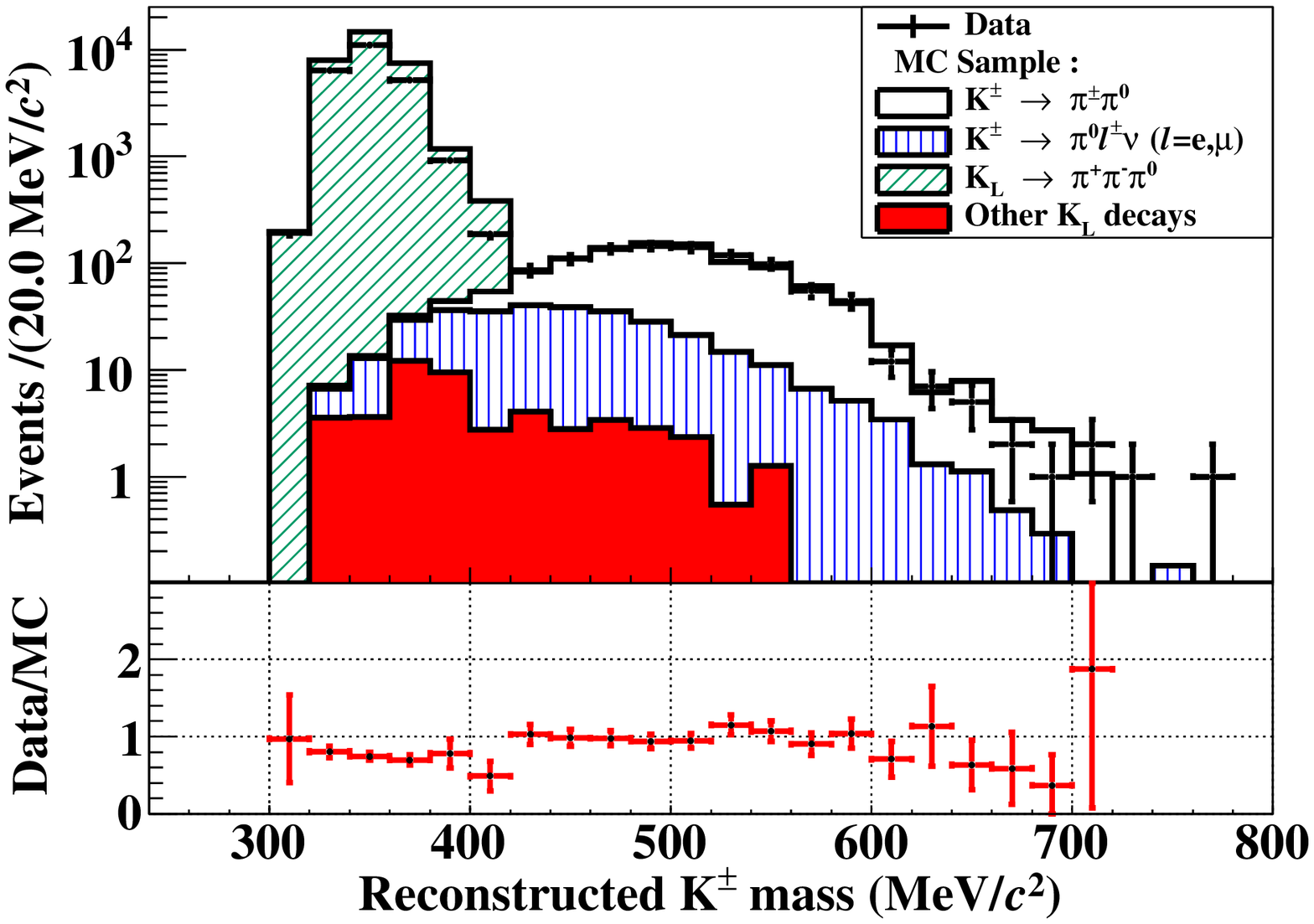}
		\caption{	
				Reconstructed $K^{\pm}$ mass ($M_{K^{\pm}}$) distribution after imposing the $\kcpipio$ selection criteria except for the requirement on $M_{K^{\pm}}$. 
				The bottom panel shows the ratio of data and MC simulations for each histogram bin.
				}
		\label{fig:KCMass}
	\end{center}
\end{figure}

\begin{figure*}
	\includegraphics[width=0.9\linewidth]{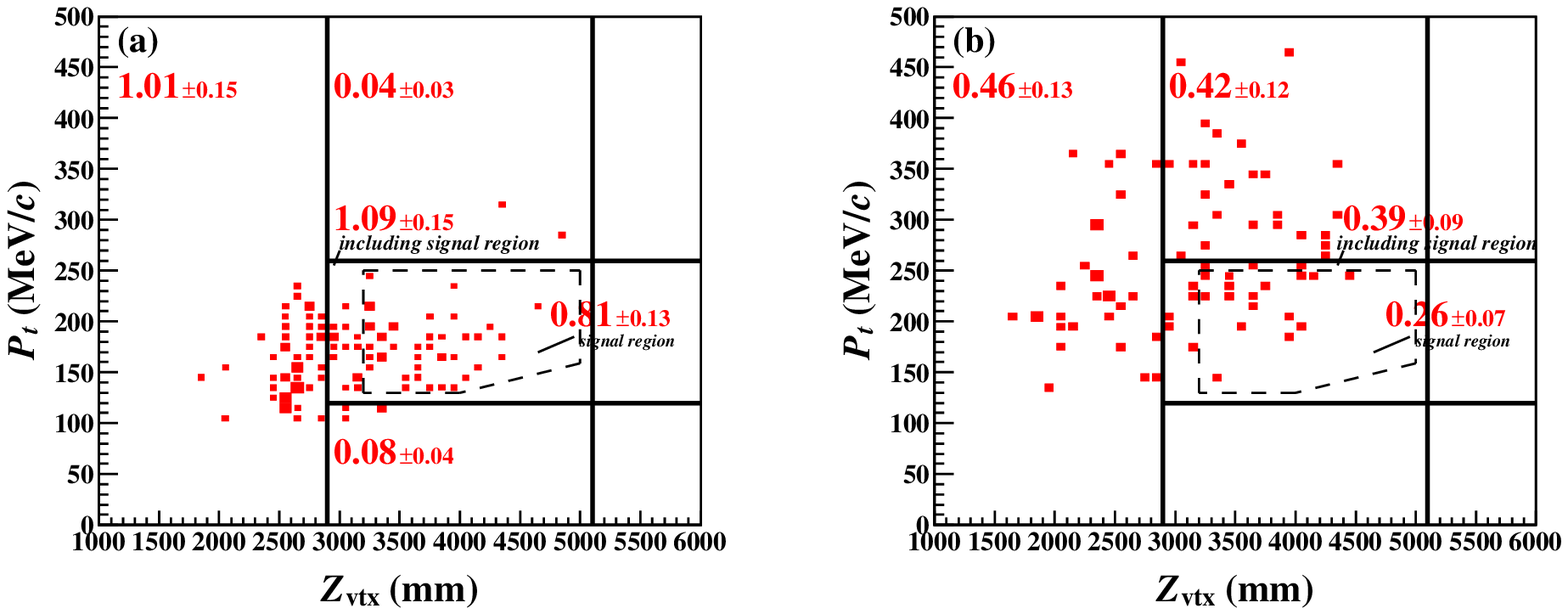}
	\caption{ 	
				Reconstructed $\pi^0$ transverse momentum ($P_t$) versus $\pi^0$ decay vertex position ($Z_{\rm vtx}$) plot of the events 
				after imposing the $\klpionn$ selection criteria on the (a) $\kcpienu$ and (b) beam-halo $\klgg$ MC simulation.
				The size of rectangles represents the number of events in arbitrary units.
				The numbers indicate the number of background events in different regions.
			}
	\label{fig:PtZ_BG}	
\end{figure*}

A $K^{\pm}$ generated in the collision of a $K_L$ with the downstream collimator can enter the KOTO detector. 
Among $K^{\pm}$ decays, $\kcpienu$ is the most likely source of background because the kinematics of the $\pi^0$ is similar to the one from the $\klpionn$ decay. 
The $K^{\pm}$ flux at the beam exit was evaluated using a $\kcpipio$ decay sample taken in 2020 with a dedicated trigger (${\pi^{\pm} \pi^0}$ trigger). 
The ${\pi^{\pm} \pi^0}$ trigger selected events with three clusters in CSI, one coincident hit in CV, and no coincident hits in other veto counters. 
In the off-line analysis, the cluster closest to the extrapolated position of the CV hit into CSI was identified as charged, while the others as neutral. 
The $\Zvtx$ was reconstructed from the two neutral clusters with the $\pi^0$ assumption. 
The $\pi^{\pm}$ direction was calculated from the $\Zvtx$ and the charged cluster position in CSI, 
and its absolute momentum was obtained by assuming the $\Pt$ balance between the $\pi^0$ and $\pi^{\pm}$. 
The energy of the charged cluster ($E_{\pi^{\pm}}$) was required to be $200\! <  E_{\pi^{\pm}} <\! 400$ MeV to select a minimum-ionizing particle. 
The reconstructed $K^{\pm}$ invariant mass ($M_{K^{\pm}}$) was required to be $440\! < M_{K^{\pm}} <\! 600$ MeV/$c^2$. 
Figure~\ref{fig:KCMass} shows the $M_{K^{\pm}}$ distribution after imposing the $\kcpipio$ selection criteria except for the requirement on $M_{K^{\pm}}$. 
Based on 847 $\kcpipio$ candidate events, the ratio of the $K^{\pm}$ to $K_L$ flux at the beam exit was measured to be $(2.6 \pm 0.1) \times 10^{-5}$. 
Figure~\ref{fig:PtZ_BG}(a) shows the $\Pt$ versus $\Zvtx$ plot of the background events from the $\kcpienu$ decay MC simulation after imposing the cuts. 
The number of background events from $K^\pm$ decays ($N_{\rm BG}^{K^\pm}$) was estimated to be $0.84 \pm 0.13$, where 97\%  comes from $\kcpienu$ decays. 
The discrepancy in the acceptance between data and MC for the cuts used in the $\klpionn$ analysis against $K^{\pm}$ decays was studied using another control sample collected in the 2020 special run. 
This control sample consisted of data taken with the physics trigger while the sweeping magnet in the beam line was turned off to enhance the $K^\pm$ flux at the beam exit. 
We simultaneously collected data with the ${\pi^{\pm} \pi^0}$ trigger in this magnet-off configuration to normalize the $K^\pm$ yield. 
We observed 27 events in the signal region after imposing the cuts to the control sample. 
This number agreed with $26.0\pm3.2$ events expected from the $K^{\pm}$ decay MC simulation. 
The ratio of these two numbers ($R_{A_{K^{\pm}}}$) was calculated to be $1.04 \pm 0.26$, where the uncertainty comes from the $K^\pm$ spectrum difference between the configurations of the magnet on and off, as well as statistical uncertainties. 
Finally, $N_{\rm BG}^{K^\pm}$ was corrected with $R_{A_{K^{\pm}}}$ and was estimated to be $0.87 \pm 0.13_{\rm stat} \pm 0.21_{\rm syst}$. 

$\klgg$ decays that occur off the beam axis can be a background source since the reconstructed $\Pt$ can be large and the cut on the projection angle no longer works. 
The yield of the beam-halo $K_L$ was evaluated by using $\klpiopiopio$ events with large $R_{\mathrm{COE}}$ values. 
After multiplying the MC expectations by the measured beam-halo $K_L$ yield, the number of the beam-halo $\klgg$ background events was estimated to be $0.26 \pm 0.06_{\rm stat} \pm 0.02_{\rm syst}$, where the systematic uncertainty comes from the MC reproducibility of the beam-halo $K_L$ spectrum. 
Figure~\ref{fig:PtZ_BG}(b) shows the $\Pt$ versus $\Zvtx$ plot of the beam-halo $\klgg$ background events from the MC simulation after imposing the cuts. 

\paragraph*{Conclusions and prospects.---}\hspace{-14pt}
With the 2016--2018 dataset, we obtained an SES of $ ( 7.20 \pm 0.05_{\rm stat} \pm 0.66_{\rm syst} ) \times 10^{-10}$ and observed three events in the signal region. 
We estimated the total number of background events to be $1.22 \pm 0.26$ with the two new background sources. 
The corresponding probability of observing three events is 13\%. 
We conclude that the number of observed events is statistically consistent with the background expectation estimated after finding two new sources. 
Assuming Poisson statistics and considering uncertainties \cite{UpperLimit}, we set an upper limit on the branching fraction of the $\klpionn$ decay in this dataset to be $4.9\times10^{-9}$ at the 90\% C.L. 

To suppress the background from $K^{\pm}$ decays in future datasets, we are preparing a new charged-particle veto counter to be installed in the beam at the upstream edge of the KOTO detector. 
We have developed and installed a prototype consisting of 1 mm$^2$ scintillation fibers in 2020 and its performance was checked. 
We are also considering to install a new sweeping magnet at the beam exit to reduce the number of $K^{\pm}$'s entering the KOTO detector. 
To suppress the background from beam halo $\klgg$ decays in future datasets, we are developing new cuts to extract the true incident angle of the photons based on the cluster energy and shape. 
We expect that these improvements will suppress backgrounds from $K^{\pm}$ and beam-halo $\klgg$ decays.

\begin{acknowledgments}	
	We would like to express our gratitude to all members of the J-PARC Accelerator and Hadron Experimental Facility groups for their support. 
	We also thank the KEK Computing Research Center for KEKCC and the National Institute of Information for SINET4. 
	This material is based upon work supported by the Ministry of Education, Culture, Sports, Science, and Technology (MEXT) of Japan and 
	the Japan Society for the Promotion of Science (JSPS) under the MEXT KAKENHI Grant No. JP18071006, 
	the JSPS KAKENHI Grants No. JP16H06343, No. JP23224007, No. JP16H02184, No. JP23654087, No. 17K05479, 
	No. JP20K14488, No. JP17J02178, and No. JP17J05397,
	through the Japan-U.S. Cooperative Research Program in High Energy Physics; 
	the U.S. Department of Energy, Office of Science, Office of High Energy Physics, 
	under Awards No. DE-SC0006497, DE-SC0007859, and No. DE-SC0009798; 
	the Ministry of Education and the Ministry of Science and Technology in Taiwan 
	under Grants No. 104-2112-M-002-021, No. 105-2112-M-002-013, and No. 106-2112-M-002-016; 
	and the National Research Foundation of Korea (Grants No. 2019R1A2C1084552 and No. 2018R1A5A1025563). 
	Some of the authors were supported by Grants-in-Aid for JSPS Fellows.
\end{acknowledgments}	


\end{document}